# Particle-Settling in Iper-viscous Liquids and its Use for Time Measurement

Gianfranco Carotenuto

*Abstract* — We live in a substantially non-linear world. Phenomena, unless approximately treated, always follow non-linear laws. Linearity is a rarity with only few exceptions (e.g., the pseudo zero-order kinetics of solid phases dissolution in liquids). A phenomenon can be used to measure time only in the case it is linear with respect to the time, that is, it occurs at a constant speed. If a body moves at a constant speed (i.e., without any acceleration), the measurement of its displacement can be used as timekeeping, in fact it is: s(t)=s₀+v·t, that is, s(t) is a linear time function. The fall of a body in a fluid is a non-linear process, since there is first an acceleration step and then a constant speed step. Therefore, in principle, it cannot be used for time measurements otherwise it leads to wrong results. However, the duration of the acceleration step reduces as the viscosity of the fluid increases, and therefore in an extremely viscous fluid there is no body acceleration, but only a very slow movement at constant speed. Owing to the linearity of the body movement, its use for timekeeping becomes very precise. Discontinuous rubbery-pitch/metal composites are characterized by the de-mixing phenomenon. It takes place because of the settling of the metallic filler dispersed into the rubbery matrix (at room temperature pitches are a kind of extremely high-viscous fluid), as the effect of forces acting on the particles (i.e., weight-force, buoyancy force and drag force). In the case of monodispersed systems, metallic particles move together and therefore there is a well-defined separation line between the pure-pitch phase and the composite phase, that continuously moves in the bottom direction as a result of the forces acting on each particle. The study of the particle falling movement and in particular of the deceleration step (named transitory stage) shows that, owing to the high viscosity of the dispersing medium, in these systems settling takes place linearly during the time. Acceleration decreases exponentially with the increasing of viscosity, while the velocity decreases linearly. Therefore, particle-settling takes place very slowly and at constant rate in liquid media with high viscosity. Such a type of composites can be industrially exploited for many functional applications like, for example, timers for very long time periods (centuries).

## I. INTRODUCTION

The gravitational field generates an attractive force on all bodies located on the earth surface. This force, named weigh-force, $\vec{p}$, is directed perpendicularly to the land surface and acts in the versus of the land. The weight-force has been exploited to fabricate many simple mechanical devices like: hourglass, inclined plane, pendulum, equal-arm balance, etc., however $\vec{p}$ can be also used in the operation of functional materials that may behave just like a hourglass, inclinometer, altimeter or other mechanical devices. The working principle of the ancient hourglass is particle-settling in air, which is a very low viscosity fluid. The special shape of hourglass is required in order to delay the settling process of the contained sand. The same principle (i.e., particle-settling) can be used for developing a material capable to behave like a hourglass. If the viscosity of the liquid medium, where particles are dispersed, is much higher than air viscosity, to delay the settling process by a special shape of the container (inclined walls to form a sort of funnel) is not necessary. Such a type of liquid dispersing media can be found among the 'apparent solids', that are extremely viscous liquids, that resemble soft solid substances. Particle-settling takes place in all fluids and, depending on the fluid viscosity value, it can be a fast, slow or extremely slow process [1,2]. In highly viscous liquid matrices in the rubbery state, like pitches slightly above the $T_g$ value, particle-settling is an extremely slow process potentially useful for time measurement. In particular, the glassy-state is characterized by a unique physical property, consisting in a temperature threshold for molecular mobility [3,4]. Crossing this temperature threshold, the molecules contained in such type of solid phases have the possibility to switch from a stationary state to a flowing condition. This special temperature is called glass-liquid transition temperature and it is denoted by the $T_g$ symbol. At temperatures higher than $T_g$ the material behaves like a liquid, with a viscosity value depending on how far from $T_g$ the temperature is. Below $T_g$, the molecules have a very limited mobility: they do not translate but can only vibrate and small functional groups can also rotate. However, frequently $T_g$ has not a single value but extends over a temperature range centred on a temperature known as "softening point" [5]. Inside the temperature interval of glass-liquid transition there are different degree of molecular mobility [6]. Molecular mobility starts at the lower interval bound and it is completed at the upper bound, while most molecules are capable to move by diffusion starting from the inflection point. When the material temperature falls inside this transient domain, the material still looks like a solid but a certain molecular mobility by diffusion becomes possible. For colophony this temperature range extends from 25°C to 60°C and it is centred on 45°C [5]-[7]. The phenomenon of glass-liquid transition in polymers has been exploited in the fabrication of different functional materials useful for industrial applications [8], and it has the potentiality to be used for developing also continuously changing materials, sensible to all force fields like for example the gravitational

G. Carotenuto, Institute for Polymers, Composites and Biomaterials (IPCB-CNR), National Research Council.
Piazzale E. Fermi, 1 – 80055 Portici (NA). Italy.
(e-mail: giancaro@unina.it)

field, that are useful for time measurement.

The possibility to use particle-settling to measure time has been described in the literature (e.g., the Stokes falling sphere clock [9]). However, the role of the dispersing medium viscosity in the precision of the time measurement and the possibility to develop functional materials and devices on this principle, to the best of our knowledge have never been described. In particular, here particle-settling in rubbery matrices has been hypothesized as working principle of self-changing materials for time measurement and theoretically described by using Newtonian fluid considerations. In general, apparent solids is a class of substances that can represent a great resource in industrial technologies, since they allow to extend to solid materials those properties belonging only to liquids like the possibility for a particle to move through them under the effect of a force (e.g., the weight-force). In particular, a model for particle-settling and the consequent possibility of designing functional devices based on particle-settling, has been developed.

## II. RESULTS

### A. Fluid-dynamic model for particle-settling in a rubbery matrix

A composite material made of perfectly spherical and monodispersed metal particles (e.g., noble metal particles, that are characterized by very high density values), uniformly dispersed into a rubbery matrix like a pitch (e.g., colophony), at a convenient temperature, that is close to the own softening point, is an unstable system spontaneously changing over long time periods. Owing to the weight-force acting on each particle of the system and the liquid nature of the dispersing medium, a settling process slowly takes place, thus allowing the system to stabilize (i.e., the potential energy lowers). The particulate matter flux through a surface, S, is defined as the number of particles crossing the surface unit in the time unit [10]. Therefore, the particle that have settled at sample bottom, n(t), (coincident with the number of particles that have cross the surface S during the time period t) is given by the following expression:

$$n(t) = S \cdot \int_0^t \varphi(t) \cdot dt \qquad (1)$$

where S is the surface area, t the time period, and $\varphi$ the particle flux. The explicit dependence of $n(t)$ on the system parameters can be calculated by considering the discrete settling of free-falling monodispersed spherical particles, moving at a velocity v(t). During a time interval, dt, the volume cross by the particle system is given by:

$$dV = S \cdot dx = S \cdot v(t) \cdot dt \qquad (2)$$

Therefore, the number of particles, dN, crossing the surface of area S is given by the following expression:

$$dN = C \cdot dV = C \cdot S \cdot v(t) \cdot dt \qquad (3)$$

where C is the suspended particle concentration (i.e., number of particles per volume unit), that has a constant value because particles are identical and move together. It is possible to write the following expression by reorganizing the terms in eq. 3:

$$\varphi = \frac{1}{S} \cdot \frac{dN}{dt} = C \cdot v(t) \qquad (4)$$

In order to derivate an expression for the speed of particles, the Newton law has to be applied to a free-falling particle subject to: weight-force, buoyancy force, and drag force (i.e., $\sum_i F_i = m \cdot \frac{dv}{dt}$). The following differential equation can be written:

$$m \cdot g - d_P \cdot V \cdot g - 3\pi \cdot D \cdot \mu \cdot v(t) = m \cdot \frac{dv(t)}{dt} \qquad (5)$$

where g is the gravity acceleration, m is the particle mass, $d_P$ is the pitch density, V the particle volume, D the particle diameter, and μ the dispersing medium viscosity. The integration of this differential equation gives the temporal law for the speed of a free-falling particle [11]:

$$v(t) = v_L \cdot \left(1 - e^{-\frac{t}{\tau}}\right) \qquad (6)$$

where τ is the time-constant, which is given by the following expression:

$$\tau = \frac{d_M \cdot D^2}{18\mu} \qquad (7)$$

and $v_L$ is the limit speed, that is given by the Stokes' law:

$$v_L = \frac{(d_M - d_P) \cdot g \cdot D^2}{18\mu} = \left(1 - \frac{d_P}{d_M}\right) \cdot g \cdot \tau \qquad (8)$$

where $d_M$ is the metal particle density. The particle acceleration temporal law, a(t), results from the derivation of eq. (6), and it is given by the following equation:

$$a(t) = \frac{v_L}{\tau} \cdot e^{-\frac{t}{\tau}} = \left(1 - \frac{d_P}{d_M}\right) \cdot g \cdot e^{-\frac{t}{\tau}} \qquad (9)$$

As a consequence, the acceleration decays exponentially with time, and its initial slope is dependent on the medium viscosity:

$$\dot{a}(t=0) = -\left[\frac{18 \cdot (d_M - d_P) \cdot g}{d_M^2 \cdot D^2}\right] \cdot \mu = -k \cdot \mu \qquad (10)$$

Therefore, the acceleration instantly cancels for very viscous fluids like pitches in their rubbery state (see Figure 1), and v can be considered as always coincident with $v_L$. This simplified expression (i.e., $v(t)=v_L$) allows to calculate the particle speed for colophony in the rubbery state (colophony is in its rubbery state at room temperature). In particular, since the metal density is higher than that of pitch (e.g., noble metals have densities close to ca. 20 g/cm³), the difference between metal and pitch densities can be approximate by the density of the metal, $d_M$. The viscosity of colophony has been measured and it is of the order of $10^7$ Pa·s at 40°C [12]. Eq. 12 can be used to estimate the particle settling speed, that resulted of 20 μm/h at 40°C for an Ir particle of 1mm (see Tab I).

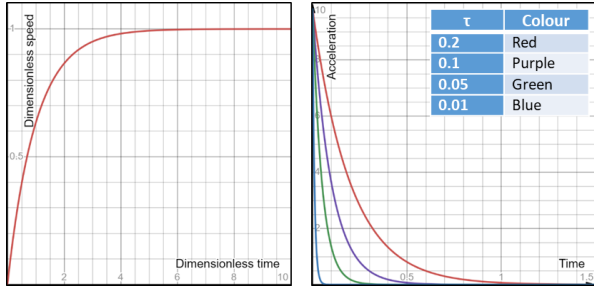

Fig. 1. Dimensionless speed behaviour with dimensionless time (left-side) and acceleration for different time-constant values (right-side).

TABLE I: PARAMETERS REQUIRED FOR THE CALCULATION OF PARTICLE SPEED IN GUM ROSIN AT 40°C.

| $d_{Iridium}$ g/cm³ | $d_{colophony}$ g/cm³ | D mm | g m/s² | η Pa·s | v μm/h |
|---|---|---|---|---|---|
| 22.56 | 1.06 | $10^{-3}$ | 9.8 | $10^7$ | 20 |

Therefore, the explicit dependence of n(t) on the system parameters is the following:

$$n(t) = S \cdot C \cdot \int_0^t (v_L - v_L \cdot e^{-\frac{t}{\tau}}) \cdot dt \quad (11)$$

$$n(t) = (S \cdot C \cdot v_L) \cdot t + \left[(S \cdot C \cdot v_L \cdot \tau) \cdot (1 - e^{-\frac{t}{\tau}})\right] \quad (12)$$

However, as above discussed, the time-constant cancels for very viscous media (i.e., τ≈0), and therefore the second term in this expression can be neglected, and n(t) reduces to:

$$n(t) = (S \cdot C \cdot V_L) \cdot t \quad (13)$$

According to equation (13), $n(t) = k \cdot t$, with $k = \frac{d_M \cdot D^2 \cdot C \cdot S \cdot g}{18\mu}$. Therefore, particles accumulate at sample bottom linearly during the time with a specific speed which grows with increasing of particle density, size, and concentration, while it decreases with increasing of liquid viscosity. Consequently, particle settling can be used for the application of time measurement because such an approach does not cause a time deformation (i.e., a progressive time expansion/contraction).

According to the expression obtained for the specific time rate, k, the calibration of this timing device (that is, the setting of an adequate value for the time flowing rate), which is required in order to have a comparability of the measurements with those achieved with other timing devices, can be easily done by modifying: metal particle size (D), type of metal ($d_M$), and particle concentration (C) in the rubbery colophony matrix. A summary of the most relevant equations that have been previously derived is given in Table II.

TABLE II: SUMMARY OF THE MOST RELEVANT EQUATIONS THAT HAVE BEEN DERIVED.

| Property | Exact formula | Apx. formula for η→∞ (τ→0) |
|---|---|---|
| Space (s) | $\left[\left(1 - \frac{d_P}{d_M}\right) \cdot g \cdot \tau^2\right] \cdot \left(\frac{t}{\tau} - 1 + e^{-\frac{t}{\tau}}\right)$ | $\left(1 - \frac{d_P}{d_M}\right) \cdot g \cdot \tau \cdot t$ |
| Velocity (v) | $\left[\left(1 - \frac{d_P}{d_M}\right) \cdot g \cdot \tau\right] \cdot \left(1 - e^{-\frac{t}{\tau}}\right)$ | $\left(1 - \frac{d_P}{d_M}\right) \cdot g \cdot \tau$ |
| Acceleration (a) | $\left(1 - \frac{d_P}{d_M}\right) \cdot g \cdot e^{-\frac{t}{\tau}}$ | 0 |

### B. Experimental methods for timing based on particle-settling

Different strategies can be potentially used to measure time by exploiting the particle-settling process. A first possibility is represented by the measurement of the electrical conductance at the sample bottom. During the particle-settling process, many close-packed particle layers stack together. For the second ohm's law, the electrical conductance, G, of the full deposited solid layer is given by the product of the single-layer conductance, $G_0$, and the number of layers (i.e., $G = n' \cdot G_0$), since all single layers have the same conductance. The number of particle layers, n', generated at sample bottom after a certain time period, can be obtained from the ratio between the number of total deposited particles at t time, n(t), and the number of particles contained inside a single layer, $n_{layer}$:

$$n' = \frac{n(t)}{n_{layer}} = \frac{k}{n_{layer}} \cdot t = K \cdot t \quad (14)$$

where $K = \frac{k}{n_{layer}}$ is a constant. Therefore, the temporal evolution of the electrical conductance at the sample bottom is given by the following function:

$$G = (K \cdot G_0) \cdot t \quad (15)$$

According to the obtained linear behaviour of the sample electrical conductance, this physical property can be conveniently used to measure the time. Electrical conductance can be very easily measured and therefore it could be really adequate for developing a time measurement device. From a practical point of view, to record a G variation on relatively short time periods, it is required for the sample to have a percolative structure, with a composition next to the inflection point of the percolation curve [13]. In fact, if the sample has a percolative structure very little displacements can determine significant changes in the material electrical conductance and therefore the self-changing property required for time measurement (i.e., conductance change) can be amplified and its change over the time becomes fast enough to be easily measured.

A further practical approach to exploit settling for time measurement is represented by the fabrication of a capacitor, that could be named: 'the Stokes capacitor'. The Stokes capacitor is a capacitor having a dielectric phase made by a dispersion of metal particles in a rubbery matrix (e.g., a pitch at a temperature slightly above the own glass-liquid transition point). The capacitor is oriented with the plates (planar or cylindrical) placed perpendicularly to the land surface. Owing to the metal particle-settling process, the capacitor is electrically equivalent to a system of two different capacitors electrically connected together in parallel (i.e., the total capacitance is the sum of the two capacitances). The first capacitor has a dielectric phase made of only pitch, while the second capacitor has a dielectric phase made of pitch-metal composite. Owing to the settling process, the particles slowly move together and settle at the device bottom, i.e., outside the dielectric field, which is present between the two plates. Therefore, the total capacitance (sum of the two capacitances), C, changes linearly during the settling process, according to the following law:

$$C = C_1 + C_2 = \varepsilon_1 \cdot \frac{S_1}{d} + \varepsilon_2 \cdot \frac{S_2}{d} = \varepsilon_1 \cdot \frac{a \cdot (b-x)}{d} + \varepsilon_2 \cdot \frac{a \cdot x}{d} = \frac{\varepsilon_1 \cdot a \cdot b}{d} - \left[\frac{(\varepsilon_1 - \varepsilon_2) \cdot a}{d}\right] \cdot x \quad (16)$$

Consequently, C is a linearly decreasing function of x and, since x is a linear function of time, x=x(t), also the device capacitance is a linear function of time.

$$C(x) = \left(\frac{\varepsilon_1 \cdot a \cdot b}{d}\right) - \left[\frac{(\varepsilon_1 - \varepsilon_2) \cdot a}{d}\right] \cdot x(t) \quad (17)$$

Also the light-scattering phenomenon can be advantageously used to monitor the particle-settling process and to exploit this process for time measurement by developing a devoted optoelectronic device.

### C. Materials with a potential energy content

Materials could be categorized as stable, unstable, and metastable, depending on how their physical/chemical properties behave over the time. Unstable materials are those material types with some physical/chemical property that changes over the time. Unstable materials are typically excluded from technological applications because of their instability of properties, however these materials can be conveniently revalorized by considering their potential use in a time measurement context. In particular, there are organic and inorganic materials that are unstable and therefore have at least one physical/chemical property that changes during the time. However, in most cases this physical/chemical property changes in a non-linear manner, and consequently they cannot be used for time measurement application. The fundamental requirement for using an unstable material for the application of time measurement is to have a monotonically changing property, that behaves linearly during the time. Indeed, time measurement can be potentially based on both chemical and physical phenomena, but to avoid time distortions and to allow comparison among time-lapse measured by different approaches a fundamental requirement must be always satisfied: the property must change linearly during the time and the growth rate should be controlled. For example, graphite oxide is a substance of unstable nature that spontaneously changes the own electrical conductance progressively over the time, going from a substantially dielectric nature to a conductive one. However, the conductance change is not linear since the graphite oxide reduction reaction to graphene has not a zero-order kinetics, and therefore it cannot be used for time measurement. Similarly, the shrinkage of a thermosetting resin filled by an electrically conductive solid phase (e.g., metallic powder) represent a further example of continuously evolving chemical process, that cannot be used for time measurement since the curing reaction has a kinetic order different from zero, and consequently the shrinkage is not evolving linearly over the time.

As a consequence, whatever type of timing device would be designed, for the comparison of measured time values, the linearity of the temporal behaviour of the physical/chemical property on which the device should operate and the possibility to modify the specific time rate of this system should be previously verified. Composite materials are adequate for fabricating unstable solids because of their high content of gravitational potential energy ( $U_{tot} = M \cdot g \cdot \frac{h}{2}$ , with M: total mass of the particles, h: sample height) coming from the presence of a filling phase that can settle at sample bottom. Composite materials made of high-density metal particles embedded in low-molecular weight amorphous substances are unstable systems above the glass-transition temperature of the organic matrix. Here, it has been verified that the settling of monodispersed spherical particles is a phenomenon that evolves linearly over the time and an expression for the specific rate has been also derived in order to establish which parameters could be modified in order to vary the specific time rate. Therefore, an unstable composite material based on the settling phenomenon can be used for time measurement. In particular, the filler settling phenomenon, can be used to develop a new electrical device exploitable for a number of industrial applications like, for example, timers for long time periods, thermal switches, solid-state chronometers, compact detonators, etc.

Commonly, particle-settling takes place in liquids, whatever is their viscosity value, but liquid materials can be hardly used for making functional devices. There are a number of very high viscous liquids, like pitches, greases, pastes, gels, etc. that seem solids, but actually behave like liquids over large time scales (relaxation). These substances can be used as matrix to fabricate self-changing composite materials. Pitches need to be in their rubbery state, that is they must be at a temperature slightly above their glass transition temperature. In order to facilitate the particle settling process, monodispersed spherical particles of high density metals, for example a noble metal like Ir, Pt, Pd, etc., can be more conveniently used. This metal-pitch composite has an electrical property at bottom and dielectric properties that are strictly depending on time. Such properties can be used as the electrical output of such novel type of smart material.

### III. CONCLUSION

Apparent solids (that is, amorphous solids like pitches, bitumen, etc.) in their rubbery state (i.e., above the glass-transition temperature, $T_g$) behave like extremely viscous liquids (iper-viscous liquids) and therefore they allow to have access to some characteristics of liquids that are potentially useful for the fabrication of new functional material classes. Particle-settling is a phenomenon typically observed for liquids, however it takes place also in low-molecular-weight rubbers like the pitch named colophony at a convenient temperature ($T_g$~45.6°C), and these materials can be used in combination with metal powders, characterized by a high density value (e.g., noble metals), to fabricate solid composite materials with self-changing physical properties (e.g., electrical conductance at bottom, dielectric constant, etc.). All timing devices operate because of the fundamental principle that a regular pattern or cycle takes place at a constant rate. According to the here presented physical model, based on the uniform settling of monodispersed spherical particles in a rubbery solid, this new material class can be exploited for time measurement because of the found linearity of the electrical property temporal evolution.